# A solution for an inverse problem in liquid AFM: calculation of three-dimensional solvation structure on a sample surface

**Ken-ich Amano**[1] **and Ohgi Takahashi**[1]

[1]*Faculty of Pharmaceutical Sciences, Tohoku Pharmaceutical University, 4-4-1 Komatsushima, Aoba-ku, Sendai, 981-8558, Japan.*

**Keywords**: Statistical mechanics of liquid, Liquid AFM, Solvation structure, Solvation force, Mean force, Contact theorem, Kirkwood superposition approximation, Convolution integral, Fourier transform.

Author to whom correspondence should be addressed: Ken-ichi Amano.

Electric mail: aman@tohoku-pharm.ac.jp

Recent frequency-modulated atomic force microscopy (FM-AFM) can measure three-dimensional force distribution between a probe and a sample surface in liquid [1]. The force distribution is, in the present circumstances, assumed to be solvation structure on the sample surface, because the force distribution and solvation structure have somewhat similar shape. However, the force distribution is exactly not the solvation structure. If we would like to obtain the solvation structure by using the liquid AFM, a method for transforming the force distribution into the solvation structure is necessary. Therefore, in this letter, we present the transforming method in a brief style. We call this method as a solution for an inverse problem, because the solvation structure is obtained at first and the force distribution is obtained next in general calculation processes.

In 2012, Amano has already proposed the other two methods for the transformation. In one method, the probe is approximately treated as an ideal probe, where its shape is same as the solvent particle and a remaining part of the probe is ignored [2,3]. The approximation is considered to be valid in a primitive stage of this study, which deepens understanding of a fundamental relation between the force distribution and solvation structure. However, the method is not accurate in all of the cases (i.e., a probe used in an experiment is not always similar to the ideal probe). In the other method, the probe is treated as a sphere with variable diameter and affinity to the solvent particles of its surface can be changed from solvophobic to solvophilic [4]. This method can change the two parameters, which are different and improved points against the former method with the ideal probe. However, its calculation condition is restricted in one-dimension. This method can be developed to the three-dimensional system, however, there is a big problem in calculation time. (It is supposed to require enormous time length when the method is used in the three-dimensional system.) The method proposed here is clearly different from the previous two methods. In the method, shape of the probe can take arbitrary shape and the system is three-dimensional. Furthermore, calculation process is completely new compared with the two previous methods.

Here, we explain a solution for the inverse problem in the liquid AFM. As introduced before, the system considered here is three-dimensional and the shape of the probe is arbitrary one. The sample surface is also able to take arbitrary shape and we set the surface as a sheet like a fixed solute. Here, we consider the probe and sheet in a simple liquid. (Simple liquid means an ensemble of small spheres. Two-body potential between two small spheres is rigid, square-well, or soft, etc. In the present letter, the simple liquid is that of a one-component fluid.) In the situation, solvation force ($f_{MP}$), mean force between the solute and sheet *minus* two-body force between them, acting on the probe can be expressed as

$$\boldsymbol{f}_{MP}(\mathbf{r}_M, \mathbf{r}_P) = \int \rho\,(\mathbf{r}; \mathbf{r}_M, \mathbf{r}_P) \frac{\partial u_{PS}(\mathbf{r}; \mathbf{r}_M, \mathbf{r}_P)}{\partial \mathbf{r}} d\mathbf{r}, \qquad (1)$$

where $\mathbf{r}_M$ and $\mathbf{r}_P$ are position vectors of the sheet (sample surface) and probe, respectively. $\rho(\mathbf{r};\mathbf{r}_M,\mathbf{r}_P)$ is number density of the solvent at $\mathbf{r}$, where the sheet and probe are positioned at $\mathbf{r}_M$ and $\mathbf{r}_P$, respectively. $u_{PS}$ is two-body potential between the probe and solvent particle. Partial differentiation of the vector $\mathbf{r}$ denotes $\partial/\partial\mathbf{r}=(\partial/\partial x)\mathbf{e}_x+(\partial/\partial y)\mathbf{e}_y+(\partial/\partial z)\mathbf{e}_z$, and $\mathbf{e}_i$ (i=$x$, $y$, or $z$) is a unit vector along $i$-axis. Eq. (1) is exactly derived from statistical mechanics of liquid by considering infinitesimal movement of the probe in the system, which justly consists with to contact theorem [5-8]. (Contact theorem explains pressure on a wall, derivation of which is performed by infinitesimal change of a system volume or solute volume.) In the AFM measurement, the sheet is fixed whereas the position of the probe is artificially changed, and they obviously do not change their own angles. In such a case, there is no need to consider rotational factors in Eq. (1). To connect the solvation force ($f_{MP}$) and solvation structure on the sheet ($g_{MS}$), we take advantage of Kirkwood superposition approximation [9] and express $\rho$ as $\rho_0 \cdot g_{MS} \cdot g_{PS}$ [10], where $\rho_0$ is bulk number density of the solvent and $g_{ij}$ is pair correlation function between i and j. Setting the position of the sheet to the origin ($\mathbf{r}_M=0$), then, $f_{MP}$ is rewritten as

$$\boldsymbol{f}_{MP}(\mathbf{r}_P) = \rho_0 \int g_{MS}(\mathbf{r}) g_{PS}(\mathbf{r} - \mathbf{r}_P) \frac{\partial u_{PS}(\mathbf{r}-\mathbf{r}_P)}{\partial \mathbf{r}} d\mathbf{r}. \qquad (2)$$

Here, we introduce a function $\boldsymbol{q}_{PS}$ which is defined as

$$\boldsymbol{q}_{\text{PS}}(\mathbf{r}-\mathbf{r}_{\text{P}}) = g_{\text{PS}}(\mathbf{r}-\mathbf{r}_{\text{P}})\frac{\partial u_{\text{PS}}(\mathbf{r}-\mathbf{r}_{\text{P}})}{\partial \mathbf{r}}. \tag{3}$$

Thus $f_{\text{MP}}$ is rewritten as

$$\boldsymbol{f}_{\text{MP}}(\boldsymbol{r}_{\text{P}}) = \rho_0 \int g_{\text{MS}}(\mathbf{r})\boldsymbol{q}_{\text{PS}}(\mathbf{r}-\mathbf{r}_{\text{P}})d\mathbf{r}. \tag{4}$$

Both sides of Eq. (4) are three-dimensional vectors so that they can be written as follows:

$$f_{\text{MP}x}(\mathbf{r}_{\text{P}})\mathbf{e}_x + f_{\text{MP}y}(\mathbf{r}_{\text{P}})\mathbf{e}_y + f_{\text{MP}z}(\mathbf{r}_{\text{P}})\mathbf{e}_z$$
$$= \rho_0 \int g_{\text{MS}}(\mathbf{r})\left[q_{\text{PS}x}(\mathbf{r}-\mathbf{r}_{\text{P}})\mathbf{e}_x + q_{\text{PS}y}(\mathbf{r}-\mathbf{r}_{\text{P}})\mathbf{e}_y + q_{\text{PS}z}(\mathbf{r}-\mathbf{r}_{\text{P}})\mathbf{e}_z\right]d\mathbf{r}. \tag{5}$$

Thus, the solvation force along $z$-axis is simply expressed as

$$f_{\text{MP}z}(\mathbf{r}_{\text{P}})$$
$$= \rho_0 \int g_{\text{MS}}(\mathbf{r}) q_{\text{PS}z}(\mathbf{r}-\mathbf{r}_{\text{P}})d\mathbf{r} = \rho_0 \int h_{\text{MS}}(\mathbf{r}) q_{\text{PS}z}(\mathbf{r}-\mathbf{r}_{\text{P}})d\mathbf{r} + \rho_0 \int q_{\text{PS}z}(\mathbf{r}-\mathbf{r}_{\text{P}})d\mathbf{r}, \tag{6}$$

where $h_{\text{MS}} = g_{\text{MS}} - 1$, which is a so-called total correlation function. The introduction of $h_{\text{MS}}$ enables us to perform the Fourier transforms, because $h_{\text{MS}}(\infty) = 0$ whereas $g_{\text{MS}}(\infty) = 1$. The second integral of the right-hand side in Eq. (6) becomes zero, because the integral represents the solvation force acting on the probe in the bulk solvent, i.e. when the probe is infinitely separated from the solid plate, the force acting on the probe is zero. (The second integral corresponds to the $z$-component of the solvation force of Eq. (2) with $g_{\text{MS}} = 1$.) Next, we define a reversal function $q^{\#}_{\text{PS}z}(\mathbf{r}_{\text{P}} - \mathbf{r}) = q_{\text{PS}z}(\mathbf{r} - \mathbf{r}_{\text{P}})$ and substitute it into Eq. (6), and then perform the forward Fourier transform on it. This operation simplifies the form of Eq. (6):

$$\tilde{f}_{\text{MP}z}(\boldsymbol{k}) = \rho_0 \tilde{h}_{\text{MS}}(\boldsymbol{k}) \tilde{q}^{\#}_{\text{PS}z}(\boldsymbol{k}), \tag{7}$$

where marks of "~" represent Fourier transforms. This equation is calculated to be

$$\tilde{h}_{\text{MS}}(\boldsymbol{k}) = \frac{\tilde{f}_{\text{MP}z}(\boldsymbol{k})}{\rho_0 \tilde{q}^{\#}_{\text{PS}z}(\boldsymbol{k})}. \tag{8}$$

Finally, our aim $g_{\text{MS}}$ can be obtained as follows:

$$g_{\mathrm{MS}}(\mathbf{r}) = \mathcal{F}^{-1}\big[\tilde{h}_{\mathrm{MS}}(\boldsymbol{k})\big] + 1, \tag{9}$$

where $\mathcal{F}^{-1}$ represents inverse Fourier transform. This is the solution for the inverse problem in the liquid AFM. This simple derivation explains how to get the solvation structure from solvation force distribution measured by the liquid AFM. In addition, it is considered that this method can be applied in an experiment of laser tweezers. Measuring a force curve between two spherical solutes by using the equipment, the solvation structure around the solute can be obtained through our theory. Concerned pointes in the method are inclusion of the Kirkwood superposition approximation and uncertain compatibility between input data of the solvation force and solvation structure around the probe. A compromise plan for the approximation is fabrication of the *modified* Kirkwood superposition approximation from theoretical or empirical standpoint. To verify this transforming method, we will make a calculation program in the near future.

**ACKNOWLEDGEMENTS**

We greatly thank Masahiro Kinoshita (Kyoto University) and Ryo Akiyama (Kyushu University) for useful advices and discussions. This work was supported by "Foundation of Advanced Technology Institute" and "Joint Usage/Research Program on Zero-Emission Energy Research, Institute of Advanced Energy, Kyoto University (ZE25B-30)".

1st Submission: 16 May (2013) EST.

2nd Submission: 11 June (2013) EST.

Two researchers were added in ACKNOWLEDGEMENTS.

3rd Submission: 21 Aug (2013) EST.

Eq. (6) was corrected, and the sentences and equations below were collaterally modified a little.

4th Submission: 27 Nov (2013) EST.

A reversal function was introduced.